\documentclass{aastex}
\usepackage{emulateapj5,apjfonts}
 
\journalinfo{The Astrophysical Journal}
 
\slugcomment{Received 2002 April 8; Accepted 2002 June 10 }

\newcommand\kms{\ifmmode {\rm km\ s}^{-1} \else km s$^{-1}$\fi} 
\newcommand\Msun{\ifmmode M_{\odot} \else $M_{\odot}$\fi} 
\newcommand\ergsec{\ifmmode {\rm ergs\ s}^{-1} \else  ergs s$^{-1}$\fi}  
\newcommand\phflux{\ifmmode {\rm photons\ s}^{-1}\;{\rm cm}^{-2} 
	\else  photons s$^{-1}$ cm$^{-2}$\fi}  
 
\shorttitle{\mbox{\boldmath $F\!U\!S\!E$} Observations of Arakelian~564} 
\shortauthors{Romano et al.\ } 
 
\received{2002 April 8}
	\begin{document} 
	\title{\mbox{\boldmath $F\!U\!S\!E$} Observations 
	of the Narrow-Line Seyfert 1 Galaxy Arakelian~564\altaffilmark{1}} 
	\author{P.\ Romano\altaffilmark{2}, 
	S.\ Mathur\altaffilmark{2}, R.\ W.\ Pogge\altaffilmark{2},
	B.\ M.\ Peterson \altaffilmark{2},   
	\&  J.\ Kuraszkiewicz\altaffilmark{3}
	}
\altaffiltext{1}{Based on observations made with the NASA--CNES--CSA {\it Far Ultraviolet
	Spectroscopic Explorer}. {\it FUSE} is operated for 
	NASA by the Johns Hopkins University under NASA contract NAS5--32985.}
\altaffiltext{2}{Department of Astronomy, The Ohio State University,  
	140 West 18th Avenue, Columbus, OH  43210--1173; 
	promano, mathur, pogge, peterson@astronomy.ohio-state.edu.}
\altaffiltext{3}{Harvard-Smithsonian Center for Astrophysics, 
60 Garden Street, MS 83 Cambridge, MA 02138; 
jkuraszkiewicz@cfa.harvard.edu.}

	\begin{abstract} 

We present a 63\,ks {\it FUSE} observation of the Narrow-Line 
Seyfert 1 galaxy \objectname[MGC +05-53-012]{Arakelian~564}. 
The spectrum is dominated by the strong emission in the 
\ion{O}{6}\,$\lambda\lambda1032, 1038$ resonance doublet. 
Strong, heavily saturated absorption troughs due 
to Lyman series of Hydrogen, \ion{O}{6} and \ion{C}{3} $\lambda 977$
at velocities near the systemic redshift of Arakelian~564 are 
also observed. 
We used the column densities of \ion{O}{6} and \ion{C}{3} in 
conjunction with the published column densities of species 
observed in the UV and X-ray bands  
to derive constraints on the physical parameters of 
the absorber through photoionization modeling. 
The available data suggest that the UV and X-ray absorbers 
in Arakelian~564 are physically related, and possibly identical.
The combination of constraints indicates that the absorber is 
characterized by a narrow range in total column density 
$N_{\rm H}$ and $U$, centered at  
$\log N_{\rm H} \approx 21$ and $\log U \approx -1.5$, 
and may be spatially extended along the line of sight.

	\end{abstract} 
 

	\section{Introduction}	

More than half of the Seyfert 1 population shows optical/UV 
intrinsic absorption associated with their active nucleus 
\citep[][ and references therein]{Crenshawea99}. 
The strong UV absorption lines, Ly$\alpha$, \ion{C}{4}, 
\ion{N}{5} (and less frequently 
\ion{Si}{4} and \ion{Mg}{2}) are found to be blueshifted, or at rest,
with respect to the narrow emission lines, providing an important indication 
of the presence of a net radial outflow of the absorbing gas. 
A similar percentage also shows an associated ionized 
(``warm'') X-ray absorber \citep{Georgeea98,Reynolds97} that is 
characterized by high ionization, $U=0.1$--$10$ 
($U=Q / 4\,\pi \, r^2 \, n_{\rm H}\,\,c$, where $Q$ is the number of 
ionizing photons)
and high total Hydrogen column density, 
$N_{\rm H} = 10^{21}$--$10^{23}$\,cm$^{-2}$,
which signature is typically the presence of \ion{O}{7} and \ion{O}{8} edges. 
During the last decade evidence has been accumulated indicating that 
the same gas is responsible for the absorption in the UV and X-ray spectra 
of Seyfert 1s 
\citep{Mathur94,MEWF94,MEW95,Crenshawea99,Krissea00,Monierea01,Brothertonea02}. 
Although it is not always possible to model the X-ray/UV absorber as a
single zone (especially when the complex UV absorption is resolved in 
multiple velocity components that are characterized by a large range 
of column densities and ionization) common characteristics of these 
absorbers have emerged, i.e., they are composed of high ionization, 
low density, high column density gas that is outflowing and is located 
in or outside the broad emission-line region (BELR). 
It is therefore worthwhile to investigate the nature of this 
nuclear component in active galactic nuclei (AGN) that represents an outflow 
(or wind) that can carry away a significant amount of kinetic energy 
at a mass-loss rate comparable to the accretion rate required to fuel the AGN
\citep{MEW95}. 

Arakelian~564 (Ark~564, IRAS 22403+2927, MGC +05-53-012) is a bright, nearby
Narrow-Line Seyfert 1 (NLS1) galaxy, with $z = 0.02467$ and  $V = 14.6$ mag
\citep{rc3.9catalogue}, and 
$L_{\mbox{\scriptsize 2--10{} keV}} = 2.4 \times 10^{43}$
\ergsec{} \citep[][hereafter Paper I]{Akn564I}. It was the object of an intense 
multiwavelength monitoring campaign that included simultaneous 
observations from 
{\it ASCA} (2000 June 1 to July 6, Paper I;  
\citealt{Poundsea01,Edelsonea02}), 
{\it HST} (2000 May 9 to July 8, \citealt{Collierea01}, Paper II; 
\citealt{Crenshawea02}, Paper IV)
and from many ground-based observatories as part of an  
AGN Watch\footnote{\anchor{http://www.astronomy.ohio-state.edu/~agnwatch}
{All publicly available data and complete references to published AGN Watch 
papers can be found at http://www.astronomy.ohio-state.edu/$\sim$agnwatch.}} 
project \citep[1998 Nov to 2001 Jan, ][ Paper III]{Shemmerea01}. 
Akn~564 has shown a strong associated UV absorber 
(\citealt{Crenshawea99}, Paper II, Paper IV). 
There are indications that it also possesses a warm X-ray absorber, 
as seen by the absorption lines of \ion{O}{7} and \ion{O}{8} detected in a 
{\it Chandra} spectrum 
\citep{Matsumotoea01}\footnote{http://www.pha.jhu.edu/groups/astro/workshop2001/papers/.}. 
	
In this paper we present the results from a $63$\,ks {\it FUSE} 
observation of Akn~564 obtained on 2001 June 29--30 UT, 
focusing in particular on the \ion{O}{6} intrinsic absorption; 
we investigate the physical properties of 
the UV and X-ray absorbing gas using the constraints on column densities 
obtained during the multiwavelength observations of this AGN. 
In \S\ref{fuseobs} we present the data. 
In \S\ref{anal} we describe our analysis methods. 
In \S\ref{photoionization} we test the hypothesis that the Warm UV-X-ray absorber 
are one and the same through photoionization calculations. 
In \S\ref{discuss} we discuss some implications of our investigation.
Our results are summarized in \S\ref{summary}. 
In a forthcoming paper (Romano et al., in preparation) we will analyze the 
intrinsic SED of Ark~564 and the properties of the gas responsible for 
the broad emission lines.

	\section{Observations and Data Reduction\label{fuseobs}}  

We observed Ark~564 with {\it FUSE} \citep{Moosea00,Sahnowea00} 
for $63$\,ks starting on 2001 June 29 07:37:42 UT. The observations, consisting  
of 24 separate exposures, were performed in photon address (time-tag) mode 
through the $30\arcsec \times 30\arcsec$ low-resolution (LWRS) aperture.
During our 

\centerline{\includegraphics[width=10.5cm,height=10.0cm]{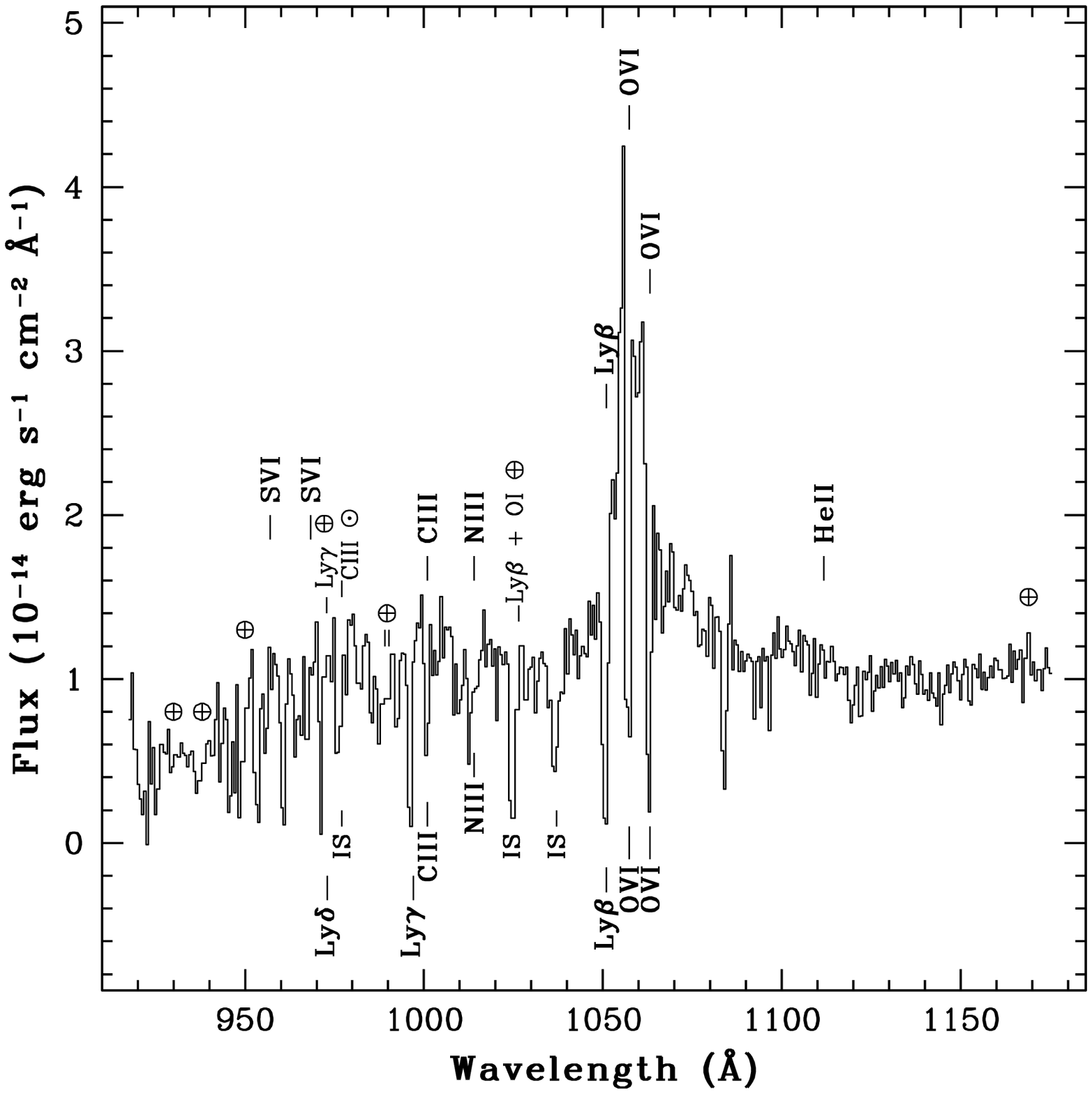}}

\figcaption{{\it FUSE} spectrum of Akn~564, 
binned to a resolution of 0.6\,\AA{} (100 pixels). In addition to the
prominent emission lines from the \ion{O}{6}\,$\lambda\lambda1032, 1038$ 
resonance doublet, 
we suggest identifications for the main emission and absorption features. 
The strong airglow lines (\ion{H}{1} Lyman series, \ion{O}{1} $\lambda$989, 
\ion{O}{1} $\lambda$1027, 
and \ion{He}{1} $\lambda$584 seen in second order at 1167\AA),  
and the \ion{C}{3} $\lambda$977 (which is scattered solar light 
in the SiC detector) have been removed. 
All other absorption features are due to Galactic or intergalactic
absorption (indicated with ``IS''). 
The absorption feature at 
$\sim 1080$\AA{} is partially due to a gap between detectors.
\label{fullspec}}

\centerline{}
\vspace{0.3cm}

\noindent
observation, a high-voltage anomaly occurred, and 
Detector 1 did not collect useful data during the first 8 exposures. 
To best model the background for this observation, 
we first used the task {\tt ttag\_combine} provided with 
the {\it FUSE} calibration pipeline, {\tt CalFUSE} 
(version 2.0.5)\footnote{See http://fuse.pha.jhu.edu/analysis/calfuse.html.}, 
to combine the last 16 exposures for Detector 1 (total exposure of 42\,ks), 
and all 24 exposures for Detector 2. 
We then processed the combined exposures with the 
standard pipeline and extracted spectra from both detectors. 
The flux scale for the final spectra is accurate to $\pm 10\,\%$,
while the wavelength scale is accurate to $\pm 15$\,\kms.
As a result of the high-voltage anomaly and data screening, the effective
on-source times were 41\,ks in Detector 1A,  39\,ks in Detector 1B,
and 58\,ks in Detector 2A, and 62\,ks in Detector 2B. 
Consequently, the SiC1A and SiC1B spectra were discarded from further analysis, 
the final signal-to-noise ratio (S/N) being $\la 1.5$ even at 0.6\AA{} 
(100 pixels) resolution.
We also discarded the LiF1B spectrum since it showed wavelength-dependent 
differences in flux of up to 30--50\,\% compared to the LiF1A 
probably due to the 
the ``worm'', which cannot be corrected for by the pipeline \citep{FUSEDHB}. 

The full {\it FUSE} spectrum was obtained by combining the spectra extracted 
from the SiC2A, LiF2B, LiF1A, SiC2B, and LiF2A segments, 
yielding a wavelength coverage of 916-1175\,\AA.
We then rebinned the full spectrum in a linear wavelength scale using 
0.07\,\AA{} bins (10 pixels, here on our full-resolution spectrum, 
with an effective resolution of 20\,\kms), 
0.2\,\AA{} bins (30 pixels, medium-resolution spectrum), and 
0.6\AA{} bins (100 pixels, low-resolution spectrum). 

Figure~\ref{fullspec} shows the low-resolution spectrum after 
we cosmetically removed 
the strong airglow lines (mainly \ion{H}{1} Lyman series, 
\ion{O}{1} $\lambda$989, 
\ion{O}{1} $\lambda$1027,  and \ion{He}{1} $\lambda$584 seen 
in second order at 1167\AA) and \ion{C}{3} $\lambda$977, 
which is scattered solar light in the SiC detector. 
The main spectral 
features are identified, the most prominent being the emission lines of the 
\ion{O}{6}\,$\lambda\lambda1032, 1038$ resonance doublet. 
Strong absorption features due to Ly$\beta$ and 
\ion{O}{6}\,$\lambda\lambda1032, 1038$
at velocities near the redshift of Ark~564 are also observed. 
We detect absorption from Galactic ISM molecular Hydrogen, 
mainly H$_2$ Lyman series absorption (see \citealt{Shullea00}, 
\citealt{Sembachea00}, and  \citealt{Savageea00}), and atomic 
Galactic ISM lines, including \ion{O}{6}\,$\lambda\lambda1032, 1038$
(Mathur et al.\ 2002, in preparation).
No intrinsic Lyman edge is detected.

	\section{Data analysis\label{anal}}	

Our goal was to determine the column densities of the ionic 
species we observed in our spectrum, combine this information with 
the column densities available in the literature for Ark~564, 
and derive constraints on the physical parameters of the absorber 
(total density and ionization) through photoionization modeling. 
{\it HST}/FOS spectra of Ark~564 \citep{Crenshawea99} show the presence 
of strong intrinsic absorption lines of Ly$\alpha$, 
\ion{N}{5}$\lambda\lambda 1238.8, 1242.8$, 
\ion{Si}{4}$\lambda\lambda 1393.8, 1402.8$, 
and \ion{C}{4}$\lambda\lambda 1548.2, 1550.8$. 
Of these lines, which are resolved in STIS spectra into 
multiple components
(Paper II, Paper IV), Ly$\alpha$, 
\ion{N}{5}, and \ion{C}{4} are completely saturated. 
Figure~\ref{o6spec} shows the {\it FUSE} full-resolution (0.07\,\AA{}, 
10 pixels) spectrum of Akn~564, in the 
Ly$\beta$/\ion{O}{6} wavelength region. 
Close examination of the \ion{O}{6} troughs
shows that the lines are heavily saturated, and their shape  is 
mainly determined by partial covering effects (see \S\ref{ALs} 
and Figure~\ref{h2normflux} below). 
Therefore, the absorption lines are not resolved into 
components at different velocities with respect to the systemic 
velocity, and we must treat each of the absorption troughs as a single 
absorption component, and we can only determine the 
velocity-averaged
column densities of the observed species. 
We note, however, that \ion{O}{6}$\lambda 1032$ is not completely black. 
Indeed, analysis of the spectra obtained with different pulse height 
restrictions and from night-only data (we did not use the latter for this work, 
since the lower S/N did not allow a proper subtraction of 
Galactic molecular Hydrogen) shows that scattered 
light is marginal in this observation, and that there is no filling 
in of the absorption troughs. 
It is also clear that the uncertainty in the measurement 
of \ion{O}{6} absorption line parameters, hence \ion{O}{6} column density,
is dominated by the uncertainty in the underlying emission-line profile. 
Additionally, there is contamination from absorption lines of Galactic 
molecular Hydrogen, with a column density log $N({\rm H_2}) \ga 16$ 
(K.\ R.\ Sembach 2002, private communication); this is not surprising, 
given the substantial amount of neutral atomic hydrogen 
($N_{\rm H} = 6.4 \times 10^{20}$ cm$^{-2}$, \citealt{DickeyL90}) 
along the line of sight toward Ark~564. 
Given these limitations, we proceeded as follows: 
we determined a power-law continuum underlying the 
Ly$\beta$/\ion{O}{6} wavelength region using the low-resolution 
spectrum (S/N $\la$\,15 in the continuum); 
we then modeled the Ly$\beta$ and \ion{O}{6} emission lines from the 
high- and medium-resolution spectra (S/N $\la 10$ in the emission 
lines for high-resolution), 
and used H$_2$ templates to estimate the H$_2$ contribution to the 
\ion{O}{6} absorption troughs. 
This part of the 

\centerline{\includegraphics[width=10.cm,height=10.0cm]{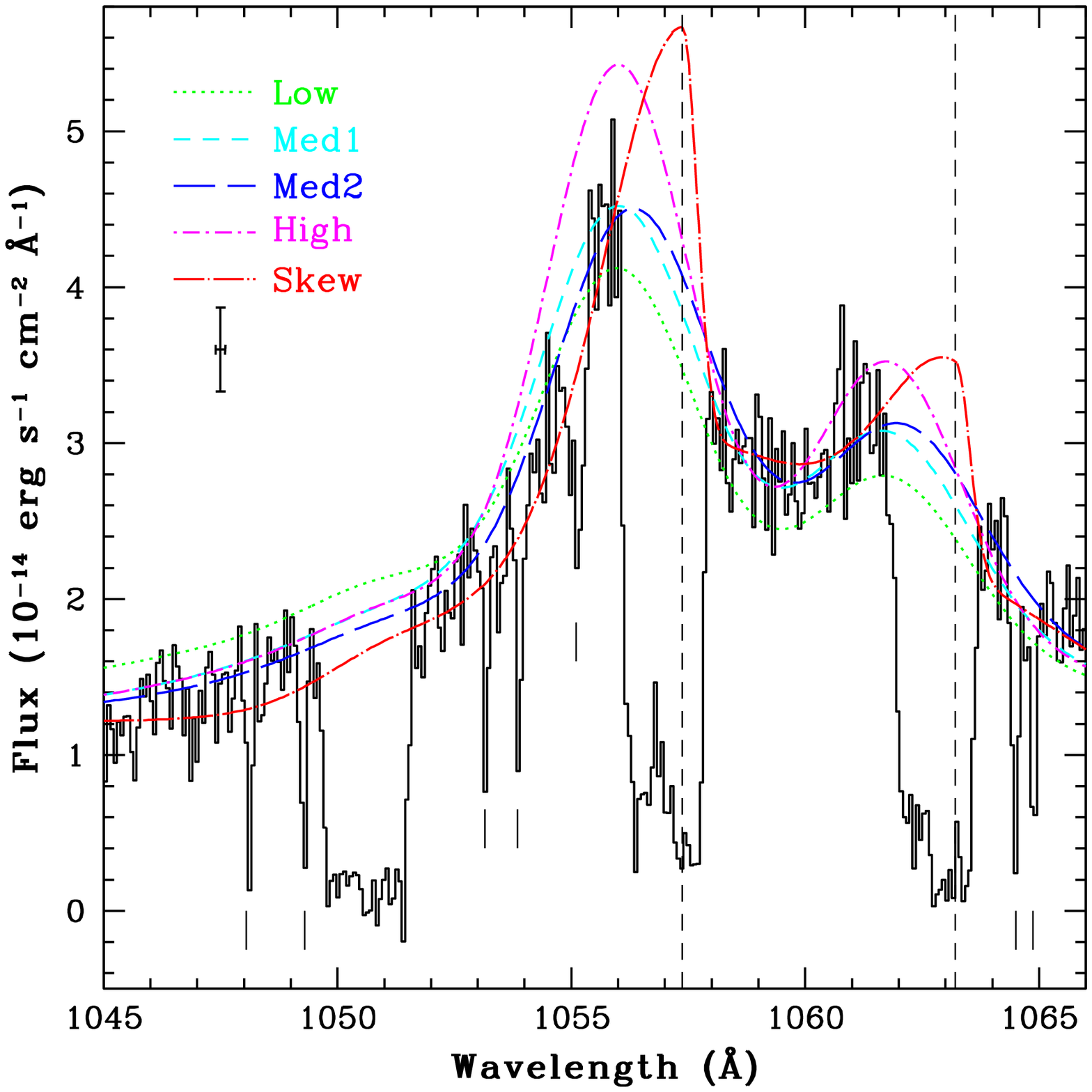}}

\figcaption{Full-resolution (0.07\,\AA{}, 
$\sim$\,10 pixels) spectrum of Akn~564, in the Ly$\beta$/\ion{O}{6} 
wavelength region. 
The dashed vertical lines mark the rest-frame wavelengths of the \ion{O}{6} lines 
and the ticks mark Galactic absorption lines. 
Overalid are our five adopted models for the 
combined continuum and emission lines, low-lying (Low), medium (Med1 and Med2),
and high-lying (High, that best follows the \ion{O}{6} $\lambda 1038$ peak), 
and a model with skewed Gaussian emission lines (Skew).
All absorption lines are saturated, but while Ly$\beta$ and 
\ion{O}{6}\,$\lambda 1038$ 
are black, \ion{O}{6}\,$\lambda 1032$ is not. The mean 1-$\sigma$ errorbar on 
the spectrum is also shown for reference. 
{\it A color version of this plot is
available in the electronic edition of this Journal.}
\label{o6spec}}

\centerline{}
\vspace{0.3cm}

\noindent
analysis was done using the {\tt IRAF}\footnote{{\tt IRAF} 
is distributed by the National Optical Astronomy Observatories, 
which are operated by the Association of Universities for Research 
in Astronomy, Inc., under cooperative agreement with the National 
Science Foundation.} task {\tt specfit} \citep{Kriss94} in the STSDAS package. 
Finally, we measured the absorption line parameters of the 
normalized line profiles using the high-resolution spectrum,  
and determined the column densities with the apparent optical depth method, 
which we briefly describe below (\S\ref{ALs}). 

	\subsection{Intrinsic \ion{O}{6} Emission Models\label{ELs}}

In addition to the power-law continuum (that we kept fixed relative to the 
fit of the low-resolution spectrum)
our models for the adopted ``continuum'' under the absorption troughs 
included
a pair of broad \ion{O}{6} emission lines (FWHM $= 4000$--5000\,\kms), 
a pair of narrow \ion{O}{6} emission lines (FWHM $= 1000$--1100\,\kms),
and a broad and narrow Ly$\beta$ emission lines. 
All lines are taken to have Gaussian profiles. 
The intensity of the \ion{O}{6} doublet lines was fixed to the 
optically thin value 2:1 for both broad and narrow lines, 
while their wavelengths were linked to the ratio of their laboratory values; 
the FWHM and wavelength of the broad Ly$\beta$ line
were linked to the ones of the \ion{O}{6} lines, though a small 
shift in wavelength was permitted, consistent with 
increasing blueshift with respect to the systemic velocity 
as the ionization increases. 

To assess the possible range of absorption-line parameters, we considered 
five different models for the emission lines, 
\centerline{\includegraphics[width=10.cm,height=10.0cm]{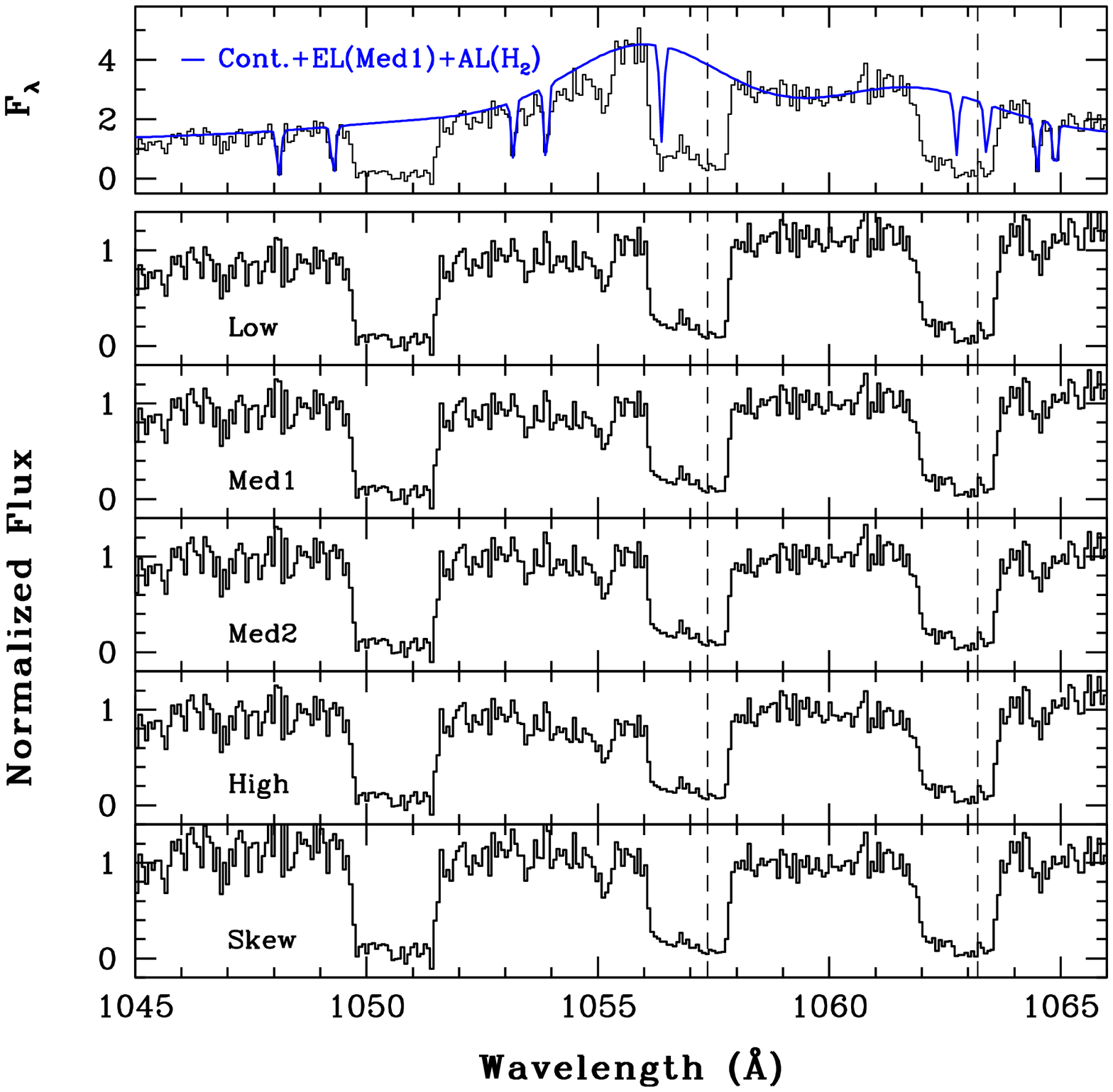}}

\figcaption{Top: 
the full-resolution spectrum of Akn~564, in the Ly$\beta$/\ion{O}{6} 
wavelength region, with a model for the continuum, the emission lines (Med1) 
and molecular Hydrogen absorption lines overlaid as a solid thicker curve
($\log N({\rm H_2}) \sim 10^{16}$).
The other panels show the normalized line profiles of the absorption 
system at $z_{\rm a} \approx z{\rm _e}$. The dashed vertical lines mark the 
rest-frame wavelengths of the \ion{O}{6} lines.
\label{h2normflux}} 
\centerline{}
\vspace{0.3cm}

\noindent
shown in Figure~\ref{o6spec}, namely, low-lying (Low), medium (Med1 and Med2), 
and high-lying (High, that best follows the \ion{O}{6} $\lambda 1038$ peak, 
though it clearly over-predicts the \ion{O}{6} $\lambda 1032$ peak). 
In all four cases the Gaussian profiles for the emission lines were 
symmetrical.
As observed in many  AGNs \citep[e.g.][]{Marzianiea96}, 
the emission lines in all our models are blueshifted 
with respect to the systemic redshift $z_{\rm e} = 0.02467$ 
as derived from \ion{H}{1} measurements 
\citep{rc3.9catalogue} by 390--1240\,\kms. 
Finally, we considered a model in which the broad lines have the least 
blueshift (100\,\kms) with respect to the systemic redshift, and the narrow 
lines are at the systemic redshift; in this case, in order to model the 
profile, the narrow emission lines must be highly asymmetrical 
(skewness $=0.2$). 
The motivation for considering the last model is the increasing evidence 
that the high-ionization emission lines in NLS1s are broader and present  
an excess of flux in the blue with respect to the low-ionization lines 
\citep[][and references therein]{Laorea97b,Petersonea00,Mathur00,Leighly01}. 
With our choice of Low, Med1, Med2, and High emission line profiles 
the absorption is redshifted with respect to the emission lines. 
This is highly unusual, though not unprecedented \citep{MEW99,Goodrich00}. 
When modeling the  emission with the Skew profile, the absorption 
is in part blueshifted, in part redshifted with respect to the 
emission lines.
Table~\ref{emissionpars} lists the \ion{O}{6} model emission-line parameters. 
Since no inflection points are clearly seen in the observed emission line 
profiles, our models may not have a direct physical interpretation. 
Also, while we believe the true shape of the emission line 
\centerline{\includegraphics[width=10.cm,height=10.0cm]{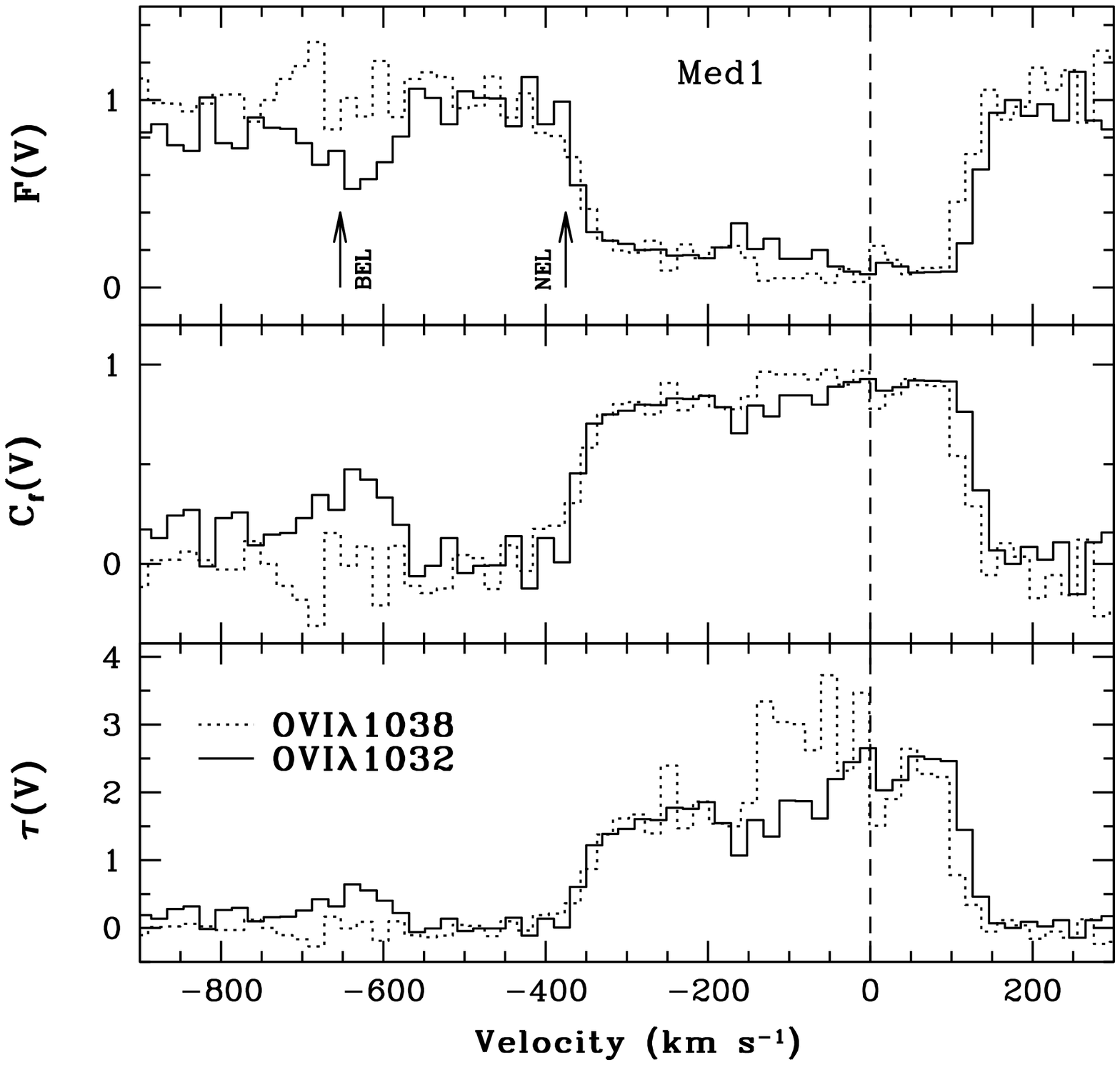}}

\figcaption{Top: normalized line profiles (Med1) of the absorption 
system as a function of radial velocity relative to a systemic 
redshift $z_{\rm e} = 0.02467$ (\ion{H}{1}).
The arrows show the position of the zero velocity with respect to the 
broad emission lines (BEL) and narrow emission lines (NEL). 
Middle: covering factor as a function of  radial velocity.
Bottom: optical depth as a function of  radial velocity.
The solid lines refers to the  \ion{O}{6} $\lambda 1032$ profile, the dotted 
line to the the \ion{O}{6} $\lambda 1038$ profile.
\label{normcftau}} 
\centerline{}
\vspace{0.3cm}

\noindent
profile may be most realistically represented by the Med1 and Med2 profiles, 
we consider the most extreme profiles (Low, High and Skew) to bracket  the 
constraints on the values of column density. 

Our spectrum also shows narrow H$_2$ and \ion{H}{1} absorption-lines 
from the inter-stellar medium (ISM) in the 
Ly$\beta$/\ion{O}{6} wavelength region; 
in particular, the top panel of Figure~\ref{h2normflux} 
shows six of such absorption lines that lie outside the absorption troughs, 
which we fit as Gaussians with {\tt specfit} (FWHM $= 30$\,\kms). 
We then used H$_2$ and \ion{H}{1} absorption-line templates 
(S.\ McCandliss 2001, private communication; K.\ R.\ Sembach 2002, 
private communication) 
to predict the position of the lines in the absorption troughs and derive 
their intensity by scaling them to the two lines at $\sim 1065$\,\AA. 
Incidentally, we note a 0.115\,\AA{} shift between the 
predicted and observed wavelengths of H$_2$ absorption lines. 
The amount of this shift is constant along our spectrum, 
hence we interpreted it as a residual zero-point offset in the 
wavelength calibration of our spectrum; 
as reported in many observations\footnote{
\anchor{http://fuse.pha.jhu.edu/analysis/calfuse\_wp1.html}{
http://fuse.pha.jhu.edu/analysis/calfuse\_wp1.html. }}, 
these offsets
can be as high as 0.25\,\AA{} in the LWRS aperture. 
We matched templates and spectra accordingly. 
Figure~\ref{h2normflux} shows the line profiles normalized with respect 
to the combined continuum, emission-line, and H$_2$ profiles, for the 
five different \ion{O}{6} emission-line models. 

	\subsection{Intrinsic \ion{O}{6} Absorption Measurements\label{ALs}} 

As mentioned above, the  \ion{O}{6} lines are so heavily 
saturated that we must treat each of the absorption troughs as a single 
absorption component (as opposed to the many components observed in 
\ion{Si}{4} and \ion{Si}{3}$\lambda 1206.5$; Paper IV).
This assumption is only strictly valid if the physical conditions are
approximately constant along the line profile, i.e., as a function of 
radial velocity. 
We consired the possibility of partial covering of the lines and used the 
apparent optical depth method to determine the column densities. 
Following \citet{Hamannea97}, we calculated the lower limit to the
line-of-sight covering factor $C_{\rm f}$ from the residual 
intensities $I_{\rm r}$ in the troughs as a function of radial velocity, 
and the corresponding apparent optical depth $\tau$, 
\begin{equation}
C_{\rm f} \geq  1 - \frac{I_{\rm r}}{I_{\rm 0}}, 
\end{equation}
\begin{equation}
\tau  \geq  \ln \, \left( \frac{I_{\rm 0}}{I_{\rm r}} \right),
\end{equation}
where $I_{\rm 0}$ is the assumed continuum intensity across the 
absorption line 
(in the case of \ion{O}{6}, $I_{\rm 0}$ is our combined power-law continuum 
and the emission line models corrected for the Galactic Hydrogen absorption 
as described  in \S\ref{ELs}).
Column densities are then obtained by integrating the apparent optical 
depth across the line profile using \citep[e.g.,][]{SS91}
\begin{equation}
N_{\rm ion}  \geq  \frac{m_{\rm e}c}{\pi e^2 f\lambda} \int \tau (v) \, dv
\end{equation} 
where $\lambda$ and $f$ are the laboratory wavelength and oscillator 
strength of the transition, respectively. 

Figure~\ref{normcftau} shows the normalized line profiles, the covering factor, 
and the optical depth as a function of  radial velocity relative to the 
systemic redshift $z_{\rm e} = 0.02467$ for the Med1 emission line profile. 
Predictably, $C_{\rm f} \approx 1$ for most of the 
\ion{O}{6} $\lambda 1038$ profile and is consistent with unity for 
\ion{O}{6} $\lambda 1032$. 
The arrows show the position of the zero velocity with respect to the 
broad emission lines (BEL) and narrow emission lines (NEL). 
As noted in \S\ref{ELs}, while most of the absorption is blueshifted 
with respect to the systemic redshift, 
the absorption troughs are completely or at least partially (Skew model) 
redshifted with respect to the broad and narrow emission lines.
The absorption troughs also show the presence of gas which is 
redshifted with respect to the systemic velocity. 
This may indicate that the absorbing gas is undergoing net radial 
infall, as is the case for NGC 5548 \citep{MEW99} and 
RX J0134-42 \citep{Goodrich00}. 

We used $\log f\lambda = 2.137$ for \ion{O}{6} $\lambda 1032$   
and  $\log f\lambda = 1.836$ for \ion{O}{6} $\lambda 1038$ \citep{Morton91}.
Table~\ref{densities} reports the values of \ion{O}{6} column densities 
for our five assumed emission line models; 
$N_{\rm O\, VI} = [2.31, 2.65] \times 10^{15}$\,cm$^{-2}$ and 
$[5.28, 5.96] \times 10^{15}$\,cm$^{-2}$ 
when measured from \ion{O}{6}~$\lambda 1032$ and 
 \ion{O}{6} $\lambda 1038$, respectively.
The errors quoted in Table~\ref{densities} are relative to the measurement 
of the integral of $\tau$ in velocity space only. 
We estimate that molecular Hydrogen lines contribute $\sim$ 10\,\% to 
the flux in the absorption troughs. 
We adopt (5.7 $\pm$ 0.07$) \times 10^{15}$ cm$^{-2}$, obtained averaging 
the values  from Med1 and Med2 emission line models for the 
\ion{O}{6} $\lambda 1038$ line, as a conservative lower limit on the 
\ion{O}{6} column density.

	\subsection{\ion{C}{3} column density}

We also determined the column density of \ion{C}{3} from the 
\ion{C}{3} $\lambda 977$ absorption trough with the apparent optical depth 
method described in \S\ref{ALs} ($\log f\lambda = 2.872$, 
\citealt{Morton91}). 
The Ly$\gamma$/\ion{C}{3} wavelength region does not necessarily 
require the same degree
\centerline{\hspace{+0.5truecm}\includegraphics[width=10.cm,height=10.0cm]{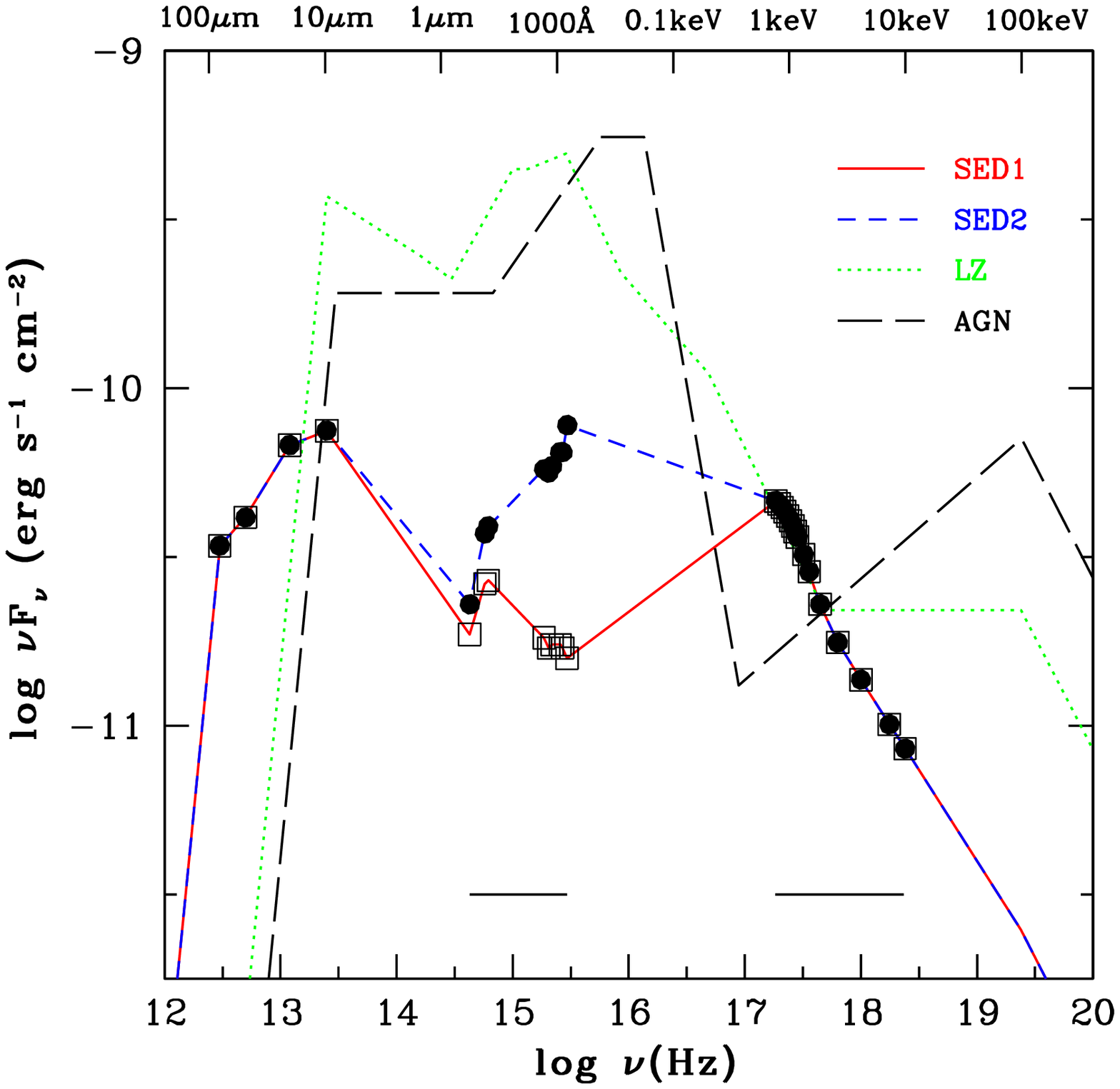}} 

\hspace{-1.truecm}

\figcaption{Comparison of the adopted SEDs for Ark~564
normalized to the absorption-corrected rest-frame flux at 2\,keV.
The long-dash line is the ``table AGN'' model in Cloudy; 
the dotted line (LZ) is the ionizing continuum described in 
	\citet{Laorea97a} and \citet{Zhengea97};
the solid line (SED1) is the Ark~564 SED one described in 
Romano et al.\ 2002 (in preparation)  
which is based on data corrected for reddening with a 
standard Galactic extinction curve and $E(B-V)=0.06$;
the short-dash line (SED2) is the Ark~564 ionizing continuum 
which is based on data corrected for reddening with both a 
standard Galactic extinction curve and $E(B-V)=0.03$ and
the intrinsic extinction curve  \citet{Crenshawea02} derive for Ark~564 
and $E(B-V)=0.14$.  The square points denote adopted 
points for SED1, the circles for SED2, while the horizontal solid lines 
show the full ranges where data were available. 
{\it A color version of this plot is
available in the electronic  edition of this Journal.}
\label{allseds}} 

\vspace{0.3cm}

\noindent
 of complication in 
the emission line profile as the 
Ly$\beta$/\ion{O}{6} wavelength region; 
however, though we performed 
fits and measurements
separate from the Ly$\beta$/\ion{O}{6} ones,  for consistency we adopted the same, 
albeit with different flux normalization, emission model as Med2: 
one broad (FWHM $= 4000$\,\kms) 
and one narrow (FWHM $= 1100$\,\kms) Gaussian emission line 
for \ion{C}{3} and one broad Ly$\gamma$ Gaussian emission line;  
the FWHM and wavelength of the broad Ly$\gamma$ line
were linked to the ones of the broad \ion{C}{3} line.
We obtain a \ion{C}{3} column density of (3.19 $\pm$ 0.05$) 
\times 10^{14}$ cm$^{-2}$
(errors are relative to the measurement of the integral of $\tau$ in 
velocity space only, while a 10\,\% contribution is due to molecular 
Hydrogen absorption lines). 

	\subsection{Velocity Centroids}

Table~\ref{constraints} reports the values of the radial velocity centroids 
relative to the systemic redshift of \ion{O}{6}, \ion{C}{3}, Ly$\beta$, and  
Ly$\gamma$ absorption lines, along with the measured column densities. 
We also show the results of Paper IV to emphasize the 
good agreement between the centroid velocity shifts (relative to systemic
redshift) obtained for the \ion{H}{1} Lyman 
series in the {\it FUSE} and {\it HST} spectra. 
In Paper IV, it is also noted that  
while saturation is probably responsible for the discrepancy in the 
values of the centroids in the different ions, the most saturated lines, i.e.,  
the \ion{H}{1} Lyman series, give us an estimate of the total 
coverage of the absorber. 
Paper IV reports a range $[-420,+180]$\,\kms{} for Ly$\alpha$, and we 
obtain $[-431,+174]$\,\kms{} for Ly$\beta$, $[-412,+147]$\,\kms{} for Ly$\gamma$,
$[-395,+177]$\,\kms{} for \ion{C}{3}, and $[-412,+130]$\,\kms{} for 
\ion{O}{6} $\lambda 1038$. 
The less saturated \ion{O}{6} $\lambda 1032$  yields $[-374,+142]$\,\kms.
Our {\it FUSE} spectrum was obtained a year after the last of the 
{\it HST}/STIS spectra were taken, and we confirm the finding of 
Paper IV  that there are no changes in radial velocity coverage 
of the absorber. 

        \section{Photoionization Modeling\label{photoionization}}

It is common practice to use photoionization codes
to predict the fractional abundance of an element in a given 
ionization state, $f_{\rm ion}$, given an input continuum, 
density $n$, total column density $N_{\rm H}$, 
and ionization parameter $U$  of the gas. 
The fractional abundance of an ion of an element X is related to its 
column density $N_{\rm ion}$ and the abundance of its element 
$N_{\rm X}$ by 
$N_{\rm ion} = N_{\rm H}\, N_{\rm X}\,N_{\rm ion}$, 
which provides a prediction of  $N_{\rm ion}$ that can be tested 
against observations. 
We used {\tt Cloudy}\footnote{http://www.pa.uky.edu/$\sim$gary/cloudy/. }
\citep[v94.00, ][]{Cloudy} 
to calculate $f_{\rm ion}$ for the ionic species for which we measured 
column densities from the {\it FUSE} spectrum, and for the species with published 
column densities, which are reported for easy reference in 
Table~\ref{constraints}. 
We considered a range of values of $N_{\rm H}$  
for a range of input continua (described in detail in \S\ref{SEDs}), 
a total Hydrogen density of $10^{5}$\,cm$^{-3}$, 
and assumed solar abundances relative to Hydrogen. 
In the case of ``table agn'', we specified a grid of $N_{\rm H}$  and 
$U$ values. 
For all other models, we normalized the SEDs  with respect to 
the measured X-ray luminosity in the absorption-corrected rest-frame
2--10\,keV energy range 
($L_{\mbox{\scriptsize 2--10{} keV}} = 2.4 \times 10^{43}$ \ergsec{}; Paper I), 
and specified the radius of the cloud, thus obtaining $U$. 
We note that the use of the observed SEDs assumes that the absorbing gas sees
the same ionizing continuum as the observer does.

	\subsection{Input continua\label{SEDs}}

Figure~\ref{allseds} illustrates our choices of input continua 
for {\tt Cloudy}. 
\begin{enumerate}
\item The {\tt Cloudy} ``table agn'' continuum, which is the  \citet{MF87}
continuum modified with a sub-millimeter break at 10\,$\mu$m, so that the 
spectral index is changed from $-1$ to $-5/2$ 
(specific flux $F_{\nu} \propto \nu^{-\alpha}$\,) for frequencies below the 
millimiter break.
While ``table agn'' is unlikely to be a representative 
spectral energy distribution (SED) of Seyferts, we use this continuum 
for comparison with the literature. 

\item A combination of the SED described in \citet{Laorea97a} and 
\citet{Zhengea97}
for radio-quiet objects (LZ in Figure~\ref{allseds}); 
we have extended the original SED in \citet{Laorea97a} 
to cover the whole $10^{-5}$--$7.354 \times 10^{6}$\, Ryd energy range 
as required by {\tt Cloudy}. 
At the low energies, we defined a sub-millimiter break and at the high energies, 
a break at 100\,keV (with a spectral index of $-5/3$),
analogous to the ones in ``table agn''. 
This ``composite'' continuum might be a typical AGN continuum. 

\item Observed SEDs: we used data obtained during the multiwavelength 
monitoring campaign performed in 2000, that included simultaneous observations 
of Ark~564 from {\it ASCA} (Paper I), {\it HST} (Paper II), 
and from many ground-based observatories (Paper III). 
In addition, {\it IRAS} measurements \citep{iras} and our {\it FUSE} 
observations have been used. 
While the full extent of 
the data is used to create a quasi-simultaneous
SED (Romano et al. in preparation), our adopted continuum for {\tt Cloudy}
only consists of selected points (also shown in Figure~\ref{allseds}). 
All data have been corrected for redshift.  

Special care has been paid in correcting the data for reddening, given the 
indications (Paper IV) that strong intrinsic neutral 
absorption is present in Ark~564 in excess of the Galactic absorption. 
Therefore, we corrected the spectra for reddening in 2 different ways: 
\begin{enumerate}
\item Using a standard Galactic extinction curve with $E(B-V)=0.06$\,mag 
	\citep{Schlegelea98} (SED1 in Figure~\ref{allseds});
\item Using a standard Galactic extinction curve with $E(B-V)=0.03$\,mag plus 
the intrinsic extinction curve \citet{Crenshawea02} (Paper IV) derive for Ark~564 
and $E(B-V)=0.14$\,mag (SED2 in Figure~\ref{allseds}). 
In the {\it FUSE} band 
we extrapolated the extinction correction linearly from the one relative to the 
{\it HST} band. 
We also note that to match our {\it FUSE} spectrum and {\it HST} spectrum in the 
overlapping region, we had to scale the {\it FUSE} fluxes by 0.75. 
This is not inconsistent with a combination of effects such as flux 
intercalibration uncertainties and, most importantly, source flux variability.
\end{enumerate}
In the X-ray, we used continuum points from the power-law fit 
(photon index $\Gamma=\alpha+1=2.538$) and added a black body component 
of temperature $T = 1.8\times 10^{6}$\,K and luminosity 
$L_{\rm bb} = 2.48 \times 10^{38}$\,\ergsec, as derived from fits to the mean 
{\it ASCA} spectrum. The square points in Figure~\ref{allseds} 
denote the adopted points for SED1, the circles for SED2, 
while the horizontal solid lines show the full ranges where data were available. 
The full details of the observed SED  as well as the analysis of the 
properties of the BELR gas will be presented in Romano et al.\ 
(in preparation). 
\end{enumerate} 

	\subsection{Physical conditions of the UV/X-ray absorber\label{absorber}}

Following \citet{Aravea01} we constrain the characteristics of the absorber
by plotting curves of constant $N_{\rm ion}$ on the $\log U$--$\log N_{\rm H}$
plane. In this plane, for each constant $N_{\rm ion}$ curve,
lower limits on the column densities, derived from
apparent optical depth line fitting methods, exclude the area below it,
while upper limits, derived from non-detections, exclude the
area above it. 
Figure~\ref{tableagn} shows the $N_{\rm ion}$ constraints 
(see Table~\ref{constraints}) for the ``table AGN'' input continuum, 
and solar abundances.  Lower limits are shown as solid lines, 
upper limits as dashed lines, detections as dotted lines. 
The combination of constraints given by the column densities suggests 
that the absorber in Ark~564 is characterized by a narrow range 
in $N_{\rm ion}$ and $U$, 
i.e., $\log U = [-1.74, -0.74]$ and  $\log N_{\rm H} = [19.90, 21.89]$. 
We note the consistency of all constraints without departure from solar 
abundances.
Analogously, Figure~\ref{mysed1} shows the  $N_{\rm ion}$ constraints for 
our SED1 input continuum and indicates 
$\log U = [-1.99, -1.31]$ 

\centerline{\vspace{-0.5cm}\hspace{+1.truecm}
	\includegraphics[width=10.cm,height=10.0cm]{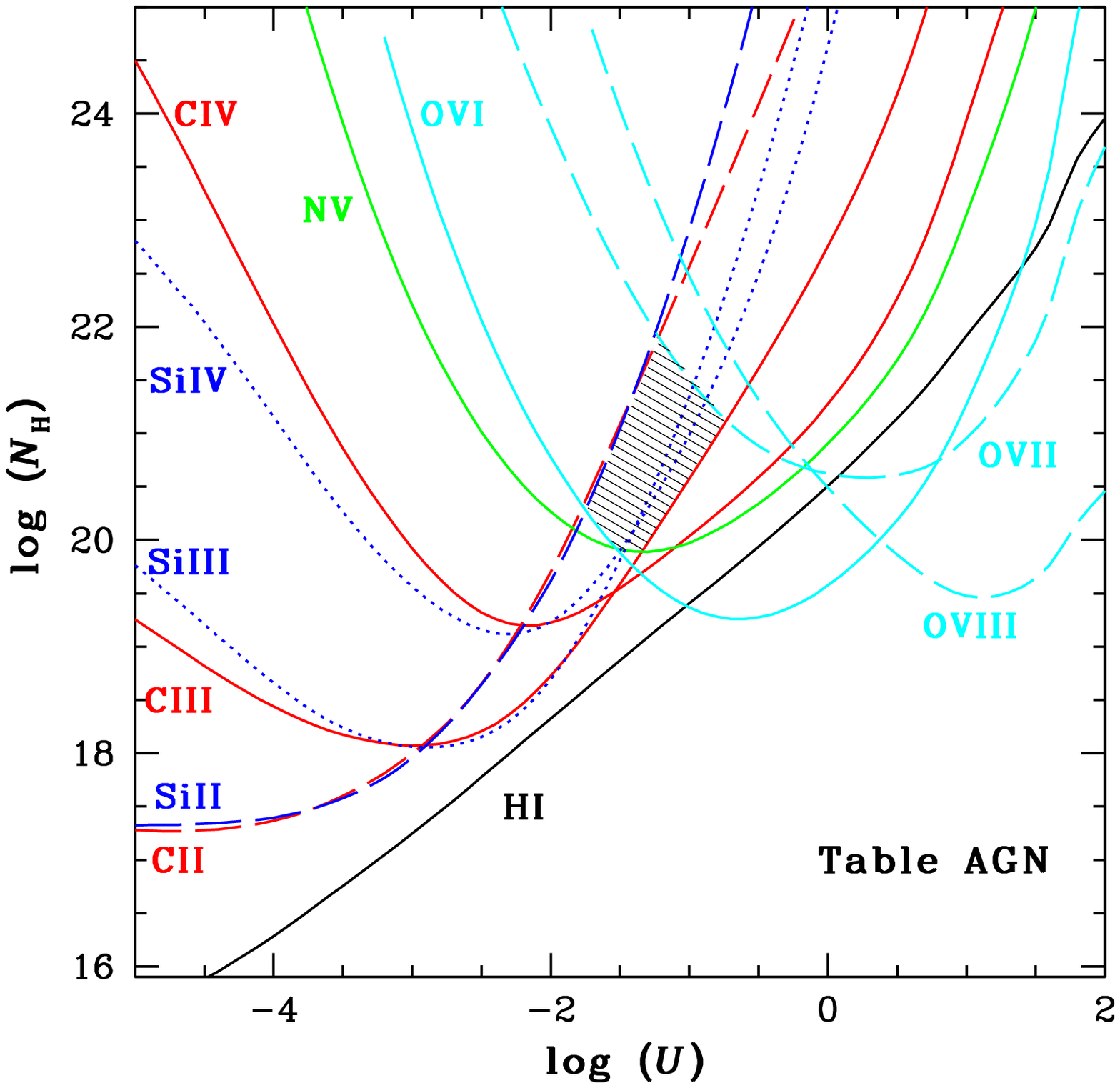}
} 

\figcaption{Photoionization curves 
at constant ionic column density on the plane of total hydrogen column density 
$N_{\rm H}$ vs.\ ionization parameter $U$. The shape of the ionizing 
radiation is defined by the ``table AGN'' model in Cloudy, and the 
abundances are solar. 
Lower limits are shown as solid lines, upper limits as dashed lines, 
detections as dotted lines. 
For clarity, curves relative to ions of the same element have been drawn 
in the same color: C in red, O in light blue, Si in dark blue. 
The shaded region corresponds to the locus on the log $U$--log $N_{\rm H}$ 
space where all conditions are met (see \S\ref{absorber}), that is, where 
$\log U = [-1.74, -0.74]$, $\log N_{\rm H} = [19.90, 21.89]$.
{\it A color version of this plot is
available in the electronic  edition of this Journal.}
\label{tableagn}}  

\centerline{\hspace{+1.truecm}\vspace{-0.5cm}
	\includegraphics[width=10.cm,height=10.0cm]{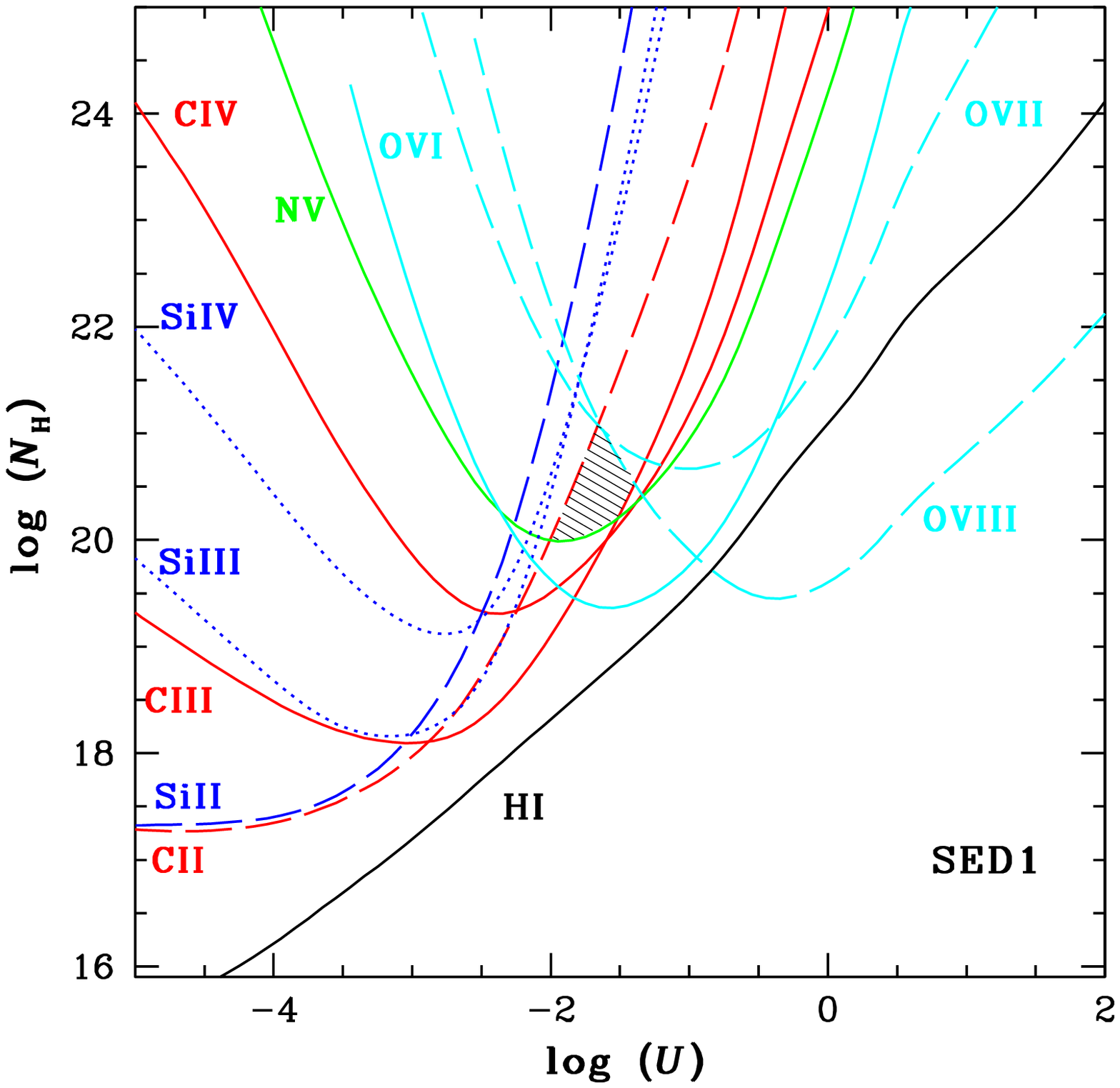}
} 
\figcaption{Same as Fig.~\ref{tableagn},
but with a ionizing continuum with no reddening intrinsic to Ark~564 
(SED1 in Figure~\ref{allseds}). Solar abundances are assumed. 
Most constraints are met in 
$\log U = [-1.99, -1.31]$, $\log N_{\rm H} = [19.99, 21.09]$ (see \S\ref{absorber}).
\label{mysed1}}

\vspace{0.6cm}

\noindent
\centerline{\vspace{-1.cm}\hspace{+1.truecm}
	\includegraphics[width=10.cm,height=10.0cm]{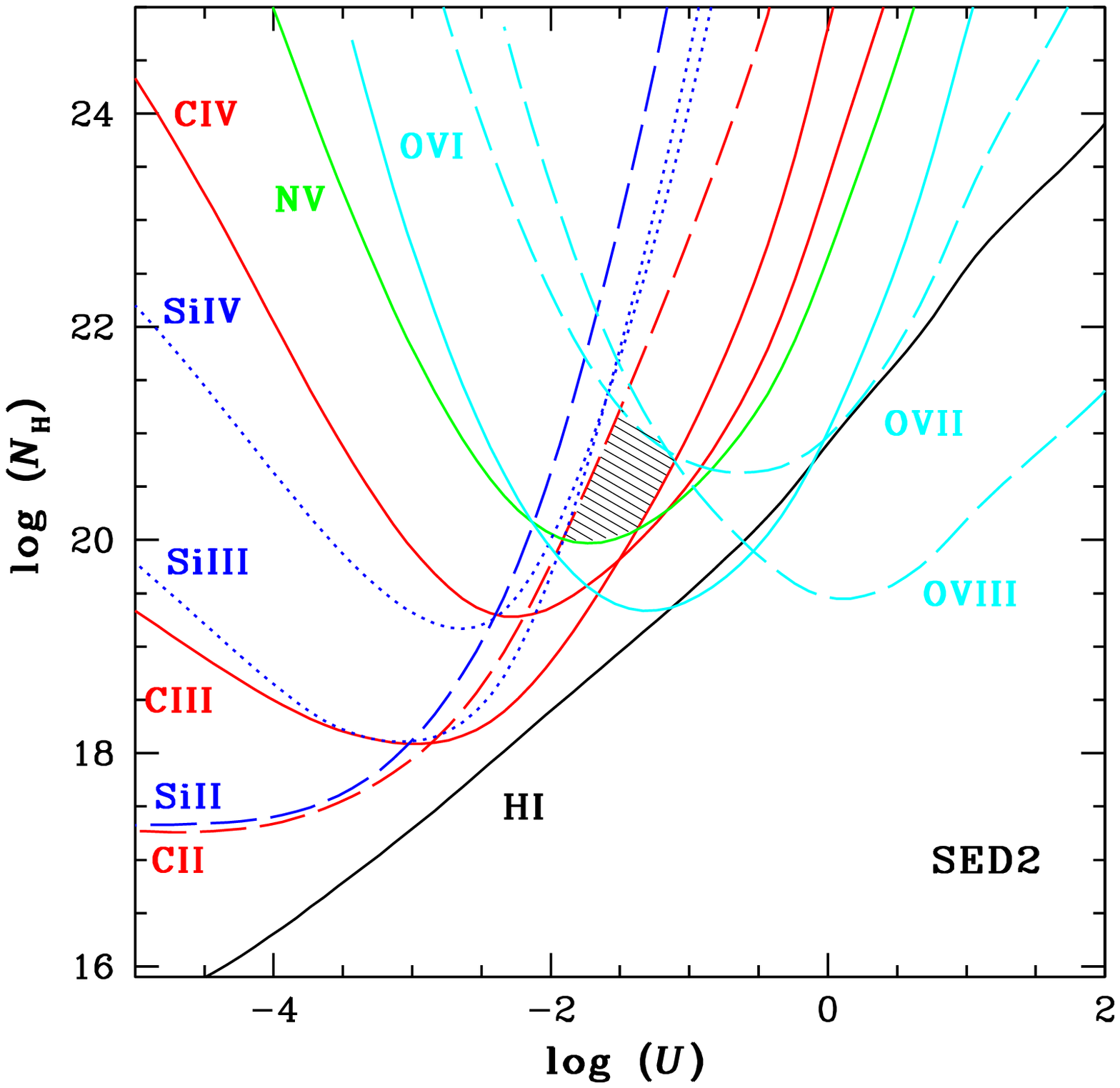}
} 
\figcaption{Same as Fig.~\ref{tableagn},
but with a ionizing continuum with Galactic and intrinsic reddening 
(SED2 in Figure~\ref{allseds}). Solar abundances are assumed. 
Most constraints are met in 
$\log U = [-1.86, -1.02]$, $\log N_{\rm H} = [19.95, 21.27]$ (see \S\ref{absorber}).
The top axis indicates the distance of the absorbing gas from the continuum source. 
\label{mysed2}} 

\vspace{0.6cm}

\noindent
and  $\log N_{\rm H} = [19.99, 21.09]$,  
 Figure~\ref{mysed2} (SED2 continuum) indicates 
$\log U = [-1.86, -1.02]$ and $\log N_{\rm H} = [19.95, 21.27]$, 
while  
Figure~\ref{lzsed} (LZ continuum) indicates 
$\log U = [-1.97, -1.54]$ and $\log N_{\rm H} = [20.00, 20.79]$.
Thus, depending on the input continuum, there are small but significant
differences in the derived properties of the absorber.

	\section{Discussion\label{discuss}}

The UV absorber in Ark~564 is in a general state of outflow with 
respect to the systemic redshift (see Table~\ref{constraints}). 
A very good agreement is found between the values of the 
velocity centroids we derive for the species observed in the {\it FUSE} 
spectrum and the ones derived for the {\it HST}/STIS spectrum 
(Paper IV); therefore, we adopt as the best estimate of the 
net radial velocity of the UV absorber the value obtained in  
Paper IV for \ion{Si}{3} and \ion{Si}{4}, the least saturated 
lines: $V_{\rm out} = -194 \pm 5$\,\kms. 
The absorption troughs also show the presence of gas which is 
redshifted with respect to the systemic velocity. 
This can be explained in part as a saturation effect, as 
is shown in Figure~3 of Paper IV.
Alternatively, a model with more than one kinematic component 
is required to explain the observed absorption troughs 
\citep[i.e.][]{Elvis2000}; in this scenario, in addition to the
blueshifted absorption from an outflowing wind, we would be observing 
redshifted absorption from infalling material, such as an accretion 
flow. 
In addition to the continuum source, the absorbing gas must cover a 
substantial portion of the BELR, since the absorption troughs 
are much deeper than the continuum level. 
Assuming the identity of the UV and X-ray absorbing gas,
we have used the column densities of the observed species 
to constrain the physical conditions of the absorber. 
For the most realistic SED (SED2), we obtained
\centerline{\vspace{-1.cm}\hspace{+1.truecm}
	\includegraphics[width=10.cm,height=10.0cm]{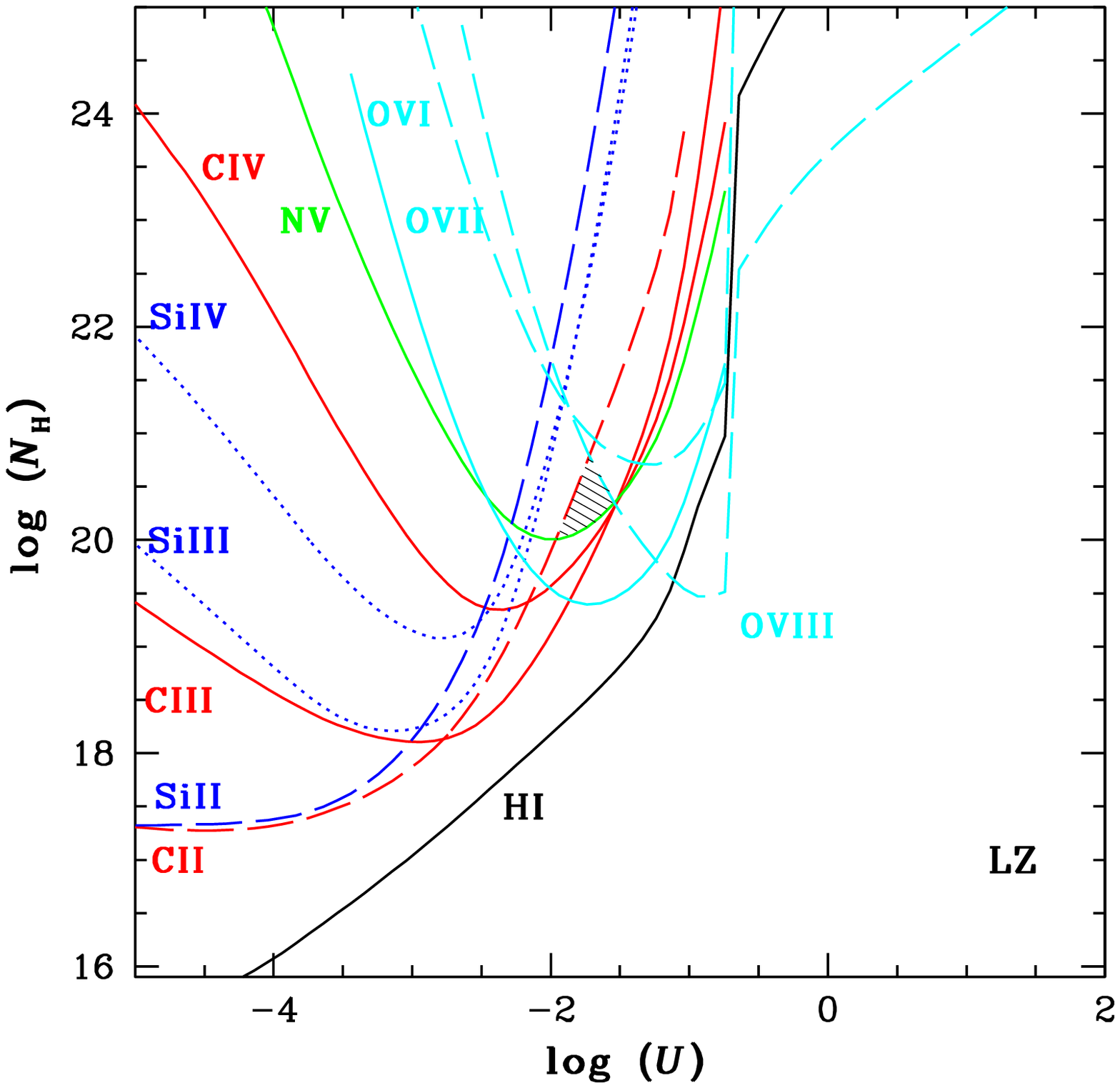}
} 
\figcaption{Same as Fig.~\ref{tableagn},
but with a ionizing continuum described in \citet{Laorea97a} and \citet{Zhengea97}
(ZL in Figure~\ref{allseds}). Solar abundances are assumed. 
Most constraints are met in 
$\log U = [-1.97, -1.54]$ and $\log N_{\rm H} = [20.00, 20.79]$
 (see \S\ref{absorber}).
\label{lzsed}} 

\vspace{0.6cm}

\noindent
$\log N_{\rm H} = [19.95, 21.27]$ and $\log U = [-1.86, -1.02]$.
These constraints can be used to determine the size of the absorber,  
its distance from the central continuum source, and the mass outflow rate. 
For the following order-of-magnitude arguments we adopt the mean values
$U = 0.0363$ and $N_{\rm H}=4.07 \times 10^{20}$\,cm$^{-2}$. 
The size of the absorber is  derived from the total column density,  
	$r_{\rm abs} = 4.07 \times 10^{20} \,n_{\rm H}^{-1}$\, cm.
For SED2 the number of ionizing photons is $Q = 6.68 \times 10^{55}$\,s$^{-1}$, 
so the distance from the continuum source is
	$R_{\rm abs} =  7.00 \times 10^{22} \, n_{\rm H}^{-1/2}$\,cm. 
Using the lower limit on $R_{\rm abs} > 95$\, pc found in Paper IV, 
this would indicate a total density $n_{\rm H} > 5.70 \times 10^{4}$\,cm$^{-3}$.
Assuming uniform density, and considering that $r_{\rm abs} \ll R_{\rm abs}$,  
the mass of the outflowing gas is 
$M_{\rm abs} = 2.11 \times 10^{10}\,\,f\,n_{\rm H}^{-1}\,\Msun$, 
where $f$ is the covering factor, i.e., the fraction of the sky covered 
by the absorber as seen at the central source. 
The mass outflow rate is then 
$\dot{M}_{\rm abs} = M_{\rm abs}\,V_{out}\,\, / r_{\rm abs} =  
3.17 \times 10^{4}\,\,f$\,\,\Msun\,yr$^{-1}$ and  
the outflow carries out a rate of kinetic energy 
$\dot{M}_{\rm abs} \,V_{\rm out}^2 / 2 = 3.76 \times 10^{44}\,f\,\,$\ergsec.
To power Ark~564 at the observed luminosity ($L_{\rm bol} = 
10 \times L_{\mbox{\scriptsize 2--10{} keV}} = 2.4 \times 10^{44}$\,\ergsec) 
at an efficiency $\eta = 0.1$, an accretion rate  
$\dot{M}_{\rm acc} = 1.8 \times 10^{-3} \left( L_{44} / \eta\right) =
4.3 \times 10^{-2}\,\Msun\,$yr$^{-1}$ is required 
($L_{44}$ is the bolometric luminosity in units of 10$^{44}$\,\ergsec).

The outflow carries out a kinetic luminosity about one order of 
magnitude smaller than the observed radiative luminosity of the source. 
However, the mass outflow rate is uncomfortably large unless the covering 
factor is very small. If $\dot{M}_{\rm abs} \la \dot{M}_{\rm acc}$, 
then it implies $f \la 10^{-6}$.
Alternatively, our assumption $r_{\rm abs} \ll R_{\rm abs}$ might not be valid. 
The absorber might be an extended, low density region.
The assumption of a uniform density gas may not be strictly valid and
the ionization parameter and density that we deduced should only be considered 
as ``average'' values. Recent {\it Chandra} observations have found extended
warm gas in some AGN \citep{Sakoea00} with physical characteristics similar
to that of a warm absorber, but seen in emission. So it is quite likely that 
the warm absorber in Ark~564 is also spatially extended along the 
line of sight. 

In Paper IV the UV absorber was modeled as a single zone with 
quasi-solar abundances (Carbon depletion being the main departure)
and the  best-fit values of  $\log U = -1.48 $ and 
$\log N_{\rm H} = 21.21$  are consistent with our limits.
\citet[][Paper IV]{Crenshawea02} over-predicted Carbon and 
Oxygen column densities:
$N_{\rm C\, III} = 5.2 \times 10^{15}$ cm$^{-2}$,
$N_{\rm O\,VI} = 2.4 \times 10^{17}$ cm$^{-2}$ 
(cf.\ our measurements: 
$N_{\rm C\, III} = 3.2 \times 10^{14}$ cm$^{-2}$ and 
$N_{\rm O\,VI} = 5.7 \times 10^{15}$ cm$^{-2}$).
These predictions, however, are consistent with the upper-end 
values of our range of parameter space. We note that our modeling did 
not require Carbon to be depleted.

Finally, we can compare our solutions of 
$\log N_{\rm H} = 21 $ and $\log U = -1.5$, 
with the preliminary results of \citet{Matsumotoea01} based on 
analysis of a 50\,ks {\it Chandra} observation of Ark~564. 
Their curve of growth analysis on the absorption lines indicates that 
$N_{\rm O\, VII}  = 3.2 \times 10^{17}$ cm$^{-2}$, 
$N_{\rm O\, VIII} = 1   \times 10^{18}$ cm$^{-2}$, 
$N_{\rm Ne\,IX}   = 3.2 \times 10^{17}$ cm$^{-2}$, and 
$N_{\rm Ne\, X}   = 1   \times 10^{17}$ cm$^{-2}$, 
suggestive of $\log N_{\rm H} = 21$ and $\log \xi = 1.6$--2. 
While there is agreement between the values of $\log N_{\rm H}$
and $N_{\rm O\, VII}$, the column densities they measure for 
\ion{O}{7} do not agree with the ones derived in Paper IV 
from the upper limits on the bound-free optical depths in the 
{\it ASCA} spectrum (Paper I). Given the high  
$N_{\rm O\, VIII}$ we would expect an edge would be observable in the 
{\it Chandra} spectrum.

	\section{Summary\label{summary}} 

We have presented a 63\,ks {\it FUSE} observation of 
the NLS1 galaxy Ark~564. The observed spectrum is dominated by the 
\ion{O}{6}\,$\lambda\lambda1032, 1038$ emission lines. 
As observed in many  AGNs \citep[e.g.][]{Marzianiea96}, 
the emission lines in all our models are blueshifted 
(or at rest, as in the case of our Skew blue-asymmetric model)
with respect to the systemic redshift by 100--1240\,\kms. 
Blue-asymmetric UV emission line profiles may be a characteristic 
of NLS1 galaxies. 

We concentrated on the analysis of the strong and heavily saturated 
absorption troughs due to Lyman series, \ion{O}{6} and 
\ion{C}{3} $\lambda 977$, which are observed at velocities near 
the systemic redshift of Ark~564. 
In a forthcoming paper (Romano et al., in preparation) we will analyze 
the intrinsic SED of Ark~564 and the properties of the BELR gas. 
Using the apparent optical depth method, we have determined that the 
column density of \ion{O}{6} is a few $10^{15}$ cm$^{-2}$, and 
$N_{\rm C\, III} = 3.2 \times 10^{14}$ cm$^{-2}$. 
We used these values in conjunction with the published column 
densities of species observed in the UV and X-ray spectra of this 
object to derive constraints on the physical parameters of 
the UV/X-ray absorbing gas through photoionization modeling. 
The combination of constraints, assuming the most realistic SED,
indicates that the absorber is characterized by a narrow range of 
density and ionization parameter, $\log N_{\rm H} = [19.95, 21.27]$
and  $\log U = [-1.86, -1.02]$. 

There is excellent agreement between the 
kinematic properties of the UV absorber emerging from the 
combined analysis of the {\it FUSE} and {\it HST}/STIS spectra, i.e.\ 
distribution of gas in radial velocity (as derived from the extent 
of the absorption troughs) and net radial velocity (as derived from 
the velocity centroids). 
The UV/X-ray absorber in Ark~564 is in outflow with respect to the 
systemic redshift with a radial velocity of a few hundred \kms, and 
it is likely spatially extended along the line of sight. 
The absorption troughs also show the presence of gas which is 
redshifted with respect to the systemic velocity. 
This may indicate that a component in the absorbing gas is 
undergoing net radial infall. This is highly unusual, though 
not unprecedented \citep{MEW99,Goodrich00}.

	\acknowledgements 

PR and SM acknowledge support through NASA grant NAG5-10320. 
We thank the {\it FUSE} team for the operation of their
satellite, and  K.\ R.\  Sembach and S.\ McCandliss for sharing their 
molecular Hydrogen absorption templates. 
We thank S.\ Vercellone for useful discussions. 
This research has made use of the NASA/IPAC Extragalactic Database
(NED) which is operated by the Jet Propulsion Laboratory, California
Institute of Technology, under contract with the National Aeronautics
and Space Administration.

\begin{deluxetable}{lllccr}	
 \tablewidth{0pc} 
 \tablecaption{\ion{O}{6} Model Emission-Line Parameters \label{emissionpars}}
\tablehead{\colhead{Model} & \colhead{Line} & \colhead{$\lambda$\tablenotemark{a}} 
		& \colhead{Flux\tablenotemark{a}} 
		& \colhead{FWHM} & \colhead{$\Delta V$\tablenotemark{b}} \\
	   \colhead{} & \colhead{} & \colhead{(\AA)} & \colhead{($10^{-13}$ erg\,s$^{-1}$\,cm$^{-2}$)}  
			& \colhead{(\kms)} & \colhead{(\kms)} \\
	   \colhead{(1)} & \colhead{(2)} & \colhead{(3)} & \colhead{(4)} & \colhead{(5)} & \colhead{(6)}
}
\startdata 
Low & BEL \ion{O}{6} $\lambda 1032$ & 1053.	& 1.00	& 5000.	& $-1240$ \\ 
 & BEL \ion{O}{6} $\lambda 1038$ & 1058.82	& 0.50	& 5000.	& $-1240$\\ 
 & NEL \ion{O}{6} $\lambda 1032$ & 1056.	& 0.90 	& 1100.	& $-390$ \\ 
 & NEL \ion{O}{6} $\lambda 1038$ & 1061.83	& 0.45 	& 1100.	& $-390$ \\ 
\hline 	
Med1 & BEL \ion{O}{6} $\lambda 1032$ & 1055.	& 1.00	& 4000.	& $-670$ \\ 
 & BEL \ion{O}{6} $\lambda 1038$ & 1060.83	& 0.50	& 4000.	& $-670$ \\ 
 & NEL \ion{O}{6} $\lambda 1032$ & 1056.	& 1.00	& 1100.	& $-390$ \\ 
 & NEL \ion{O}{6} $\lambda 1038$ & 1061.83	& 0.50	& 1100.	& $-390$\\ 
\hline	
Med2 & BEL \ion{O}{6} $\lambda 1032$ & 1056. 	& 1.00	& 4000.	& $-390$ \\ 
 & BEL \ion{O}{6} $\lambda 1038$ & 1061.83	& 0.50	& 4000.	& $-390$ \\ 
 & NEL \ion{O}{6} $\lambda 1032$ & 1056.3	& 1.00	& 1100.	& $-300$ \\ 
 & NEL \ion{O}{6} $\lambda 1038$ & 1062.13 	& 0.50	& 1100.	& $-300$ \\ 
\hline	
High & BEL \ion{O}{6} $\lambda 1032$ & 1055. 	& 1.00 	& 4000. & $-670$ \\ 
 & BEL \ion{O}{6} $\lambda 1038$ & 1060.83 	& 0.50 	& 4000. & $-670$ \\ 
 & NEL \ion{O}{6} $\lambda 1032$ & 1056. 	& 1.25 	& 1000. & $-390$ \\ 
 & NEL \ion{O}{6} $\lambda 1038$ & 1061.83 	& 0.63 	& 1000. & $-390$ \\ 
\hline	
Skew & BEL \ion{O}{6} $\lambda 1032$ & 1057.	& 1.25	& 2000.	& $-100$ \\ 
 & BEL \ion{O}{6} $\lambda 1038$ & 1062.84	& 0.63 	& 2000.	& $-100$ \\ 
 & NEL \ion{O}{6} $\lambda 1032$ & 1057.37	& 0.60	& 1000.	& 0.  \\ 
 & NEL \ion{O}{6} $\lambda 1038$ & 1063.21	& 0.30	& 1000.	& 0.  \\ 
\enddata 
\tablenotetext{a}{Observed values. }
\tablenotetext{b}{Velocities are relative to the systemic 
redshift $z_{\rm e} = 0.02467$ (\ion{H}{1} measurements, \citealt{rc3.9catalogue}).
The shift toward longer wavelengths of 0.115\,\AA{} to match 
the Galactic molecular Hydrogen  templates is not included.}
\end{deluxetable}

\begin{deluxetable}{llllll}	
 \tablewidth{0pc} 
 \tablecaption{\ion{O}{6} Column Densities from Intrinsic Absorption\label{densities}}
\tablehead{\colhead{Line} & \colhead{Low Model} & \colhead{Med1 Model} 	& \colhead{Med2 Model} 
			  & \colhead{High Model} & \colhead{Skew Model} \\
	   \colhead{} 	& \colhead{($10^{15}$ cm$^{-2}$)} 
			& \colhead{($10^{15}$ cm$^{-2}$)}  
			& \colhead{($10^{15}$ cm$^{-2}$)} 
			& \colhead{($10^{15}$ cm$^{-2}$)} 
			& \colhead{($10^{15}$ cm$^{-2}$)} \\
	   \colhead{(1)} 
			& \colhead{(2)} & \colhead{(3)} & \colhead{(4)} & \colhead{(5)} & \colhead{(6)} 
}
\startdata 
 \ion{O}{6} $\lambda 1032$ & 2.31 $\pm$ 0.03 & 2.47 $\pm$ 0.05 & 2.54 $\pm$ 0.03 & 2.65 $\pm$ 0.03 & 2.80 $\pm$ 0.03 \\ 
 \ion{O}{6} $\lambda 1038$ & 5.28 $\pm$ 0.05 & 5.60 $\pm$ 0.05 & 5.77 $\pm$ 0.05 & 5.96 $\pm$ 0.05 & 5.64 $\pm$ 0.05 \\ 
\enddata 
\tablecomments{The errors quoted are relative to the 
measurement of the integral of $\tau$ in velocity space only. 
We estimate that molecular Hydrogen lines contributes $\sim$ 10\,\% to 
the flux in the absorption troughs.}
\end{deluxetable} 

\begin{deluxetable}{llccccc}	
 \tablewidth{0pc} 
 \tablecaption{Column Densities from Intrinsic Absorption in Arakelian~564\label{constraints}}
\tablehead{\colhead{Ion} & \colhead{Wavelength/} & \colhead{Lower Limit} & \colhead{Detection} 
			& \colhead{Upper Limit} & \colhead{Velocity} & \colhead{References} \\
	   \colhead{} & \colhead{Energy} & \colhead{(cm$^{-2}$)} & \colhead{(cm$^{-2}$)}  
			& \colhead{(cm$^{-2}$)} & \colhead{(\kms)} & \colhead{} \\
	   \colhead{(1)} & \colhead{(2)} & \colhead{(3)} & \colhead{(4)} 
			& \colhead{(5)} & \colhead{(6)} & \colhead{(7)} 
}
\startdata 
 \ion{H}{1}   	& 973\,\AA{}		& \nodata		&  \nodata		& \nodata & $-108$ & 1 \\
 \ion{C}{3}    	& 977.0\,\AA{}		& $3.2 \times 10^{14}$	&  \nodata		& \nodata & $-153$ & 1 \\ 
 \ion{H}{1}   	& 1025\,\AA{}		& \nodata		&  \nodata		& \nodata & $-111$ & 1 \\
 \ion{O}{6}\tablenotemark{a} & 1031.9,1037.6\,\AA{} & $5.7 \times 10^{15}$ &  \nodata	& \nodata & $-79/-116$ & 1 \\ 
 \ion{Si}{3}   	& 1206.5\,\AA{}		& \nodata		&  $2.6 \times 10^{13}$ & \nodata & $-190$ & 2 \\
 \ion{H}{1}   	& 1216\,\AA{}		& $ 1.4 \times 10^{15}$	&  \nodata		& \nodata & $-106$ & 2 \\
 \ion{N}{5}   	& 1238.8,1242.8\,\AA{} 	& $ 3.1 \times 10^{15}$	&  \nodata		& \nodata & $-152$ & 2 \\
 \ion{Si}{2}  	& 1260.4\,\AA{}		& \nodata		&  \nodata		& $7.4 \times 10^{13}$ & \nodata & 2 \\ 
 \ion{C}{2}    	& 1334.5\,\AA{}		& \nodata		&  \nodata		& $5.4 \times 10^{13}$ & \nodata & 2 \\
 \ion{Si}{4}   	& 1393.8,1402.8\,\AA{} 	& \nodata		&  $1.6 \times 10^{14}$ & \nodata & $-197$  	& 2 \\
 \ion{C}{4}    	& 1548.2,1550.8\,\AA{} 	& $ 2.5 \times 10^{15}$	&  \nodata		& \nodata & $-130$  	& 2 \\
 \ion{O}{7}\tablenotemark{b} & 0.74\,keV& \nodata		&  \nodata		& $2.2 \times 10^{17}$ & \nodata & 2 \\ 
 \ion{O}{8}\tablenotemark{b} & 0.87\,keV& \nodata		&  \nodata		& $1.1 \times 10^{16}$ & \nodata & 2 \\ 
\enddata 
\tablecomments{Col.\ (1): Ion. 
Col.\ (2): Wavelength of absorption lines or energy of absorption edges. 
Col.\ (3): Lower limit on column densities as derived from line 
fitting/apparent optical depth methods. 
Col.\ (4): Column density values derived using the multiplet method 
of \citet{Hamannea97}.
Col.\ (5): Upper limit on column densities as derived from non-detection of the line. 
Col.\ (6): Velocity centroids relative to the systemic redshift $z_{\rm e} = 0.02467$.
Col.\ (7): References for column densities and velocity centroids. }
\tablenotetext{a}{Average of values obtained from Med1 and Med2 emission line models.}
\tablenotetext{b}{Absorption edge.}
\tablerefs{(1) This work. (2) \citet{Crenshawea02}.  
}
\end{deluxetable} 
\clearpage

\end{document}